\documentclass[a4paper,12pt,reprint,onecolumn,aip]{revtex4-1}
\usepackage[utf8]{inputenc}
\usepackage{refstyle}
\usepackage[version=3]{mhchem}

\begin{document}

\newcommand*\citeref[1]{ref. \citenum{#1}}
\newcommand*\citerefs[1]{refs. \citenum{#1}}
\newcommand*\etal{\emph{et al.}}
\newcommand*\eg{\emph{e.g.}}
\newcommand*\ie{\emph{i.e.}}
\newcommand*\LibXC{{\sc LibXC}}
\newcommand*\XCFun{{\sc XCFun}}
\newcommand*\PsiFour{{\sc Psi4}}
\newcommand*\PySCF{{\sc PySCF}}

\newcommand*\ERI[2]{\ensuremath{(#1|#2)}}
\newcommand*\dr{\ensuremath{{\rm d}\vec{r}}}
\newcommand*\dro{\ensuremath{{\rm d}\vec{r}_1}}
\newcommand*\drt{\ensuremath{{\rm d}\vec{r}_2}}

\newcommand*\trans[1]{\ensuremath{#1^\text{T}}}

\title{An overview of self-consistent field calculations within finite basis sets}

\begin{abstract}
A uniform derivation is presented of the self-consistent field
equations in a finite basis set.
Both restricted and unrestricted Hartree--Fock (HF) theory as well as
various density functional approximations are considered.
The unitary invariance of the HF and density functional models is discussed, paving the
way for the use of localized molecular orbitals.
The self-consistent field equations are derived in a non-orthogonal
basis set, and their solution is discussed in the presence of linear
dependencies in the basis set.
It is argued why iterative diagonalization of the Kohn--Sham--Fock
matrix leads to the minimization of the total energy.
Alternative methods for the solution of the self-consistent field
equations via direct minimization as well as stability analysis are
also briefly discussed.
Explicit expressions are given for the contributions to the
Kohn--Sham--Fock matrix up to meta-GGA functionals.
Range-separated hybrids and non-local correlation functionals are also
briefly discussed.
\end{abstract}

\author{Susi Lehtola}
\email{susi.lehtola@alumni.helsinki.fi}
\affiliation{Department of Chemistry, University of Helsinki, P. O. Box 55 (A. I. Virtasen aukio 1), Helsinki, Finland.}
\author{Frank Blockhuys}
\affiliation{Department of Chemistry, University of Antwerp, Groenenborgerlaan 171, 2020 Antwerpen, Belgium.}
\author{Christian Van Alsenoy}
\email{kris.vanalsenoy@uantwerpen.be}
\affiliation{Department of Chemistry, University of Antwerp, Groenenborgerlaan 171, 2020 Antwerpen, Belgium.}
\maketitle

\section{Introduction \label{sec:intro}}

Electronic structure calculations have become a cornerstone of
modern-day research in chemistry and materials physics, allowing
\emph{in silico} modeling of chemical reactions and \eg{} the first
principles design of novel catalysts.\cite{Poree2017}
Electronic structure calculations on molecular systems most often
employ the linear combination of atomic orbitals (LCAO) approach,
where the molecular orbitals (MOs) are expanded in terms of atomic
orbitals (AOs).
Several possible alternatives for the form of the AOs are commonly
used---Gaussian-type orbitals (GTOs), Slater-type orbitals (STOs), as
well as numerical atomic orbitals (NAOs); see \citeref{Lehtola2019c}
for details.
LCAO electronic structure calculations involve a variational
minimization of the total energy with respect to the AO expansion
coefficients of the MOs.
Importantly, the formalism used in the LCAO approach is not restricted
to AOs which are atom-centered basis functions; it can also be used
\eg{} in combination with numerical basis functions such as in the
finite element approach, as has been recently demonstrated in
\citerefs{Lehtola2019a} and \citenum{Lehtola2019b}.
Once the energy has been minimized and the corresponding wave function
has been obtained, it is possible to compute a number of properties
from the electronic wave function, such as the dipole moment of the
molecule and its vibrational frequencies.

The mathematical foundations for spin-restricted Hartree--Fock (HF)
theory within the LCAO approach were laid out independently by
Roothaan and Hall.\cite{Roothaan1951, Hall1951}
In their seminal papers, Roothaan and Hall derived matrix equations
that can be conveniently implemented on a computer as an iterative
procedure.
As will be seen later in \secref{scfeqs}, the Roothaan--Hall equations
turn out to yield a generalized eigenvalue problem ${\bf F C} = {\bf S
  C E}$ in the non-orthogonal AO basis set, which had been solved some
years before by Löwdin in the context of Heitler--London
theory.\cite{Lowdin1970}

Subsequently to the work by Roothaan and Hall, Pople and
Nesbet\cite{Pople1954} and Berthier\cite{Berthier1954} independently
published the corresponding equations for an unrestricted (open-shell)
HF description by an analogous scheme, without providing an explicit
derivation.
The Pople--Nesbet--Berthier equations assume a form similar to the
Roothaan--Hall equations---constituting a coupled set of general
eigenvalue equations---as will also be seen later on in the manuscript
(\secref{scfeqs}).
Restricted open-shell HF was then described by
Roothaan;\cite{Roothaan1960} restricted open-shell calculations will
not be considered in the present work as they have been extensively
reviewed by Krebs in \citeref{Krebs1999} to which we refer for further
details.

Density functional theory\cite{Hohenberg1964, Kohn1965} (DFT; see also
\citerefs{Baerends2000} and \citenum{Geerlings2003}) became popular in
chemistry through the efforts of Pople and coworkers in making the
method generally available to quantum chemists\cite{Pople1992} and
showing that atomization energies from DFT may agree well with
experiment.\cite{Johnson1993a, Johnson1994b}
Also DFT turns out to yield self-consistent field (SCF) equations that
assume the same form as in HF but with a different expression for the
Fock matrix ${\bf F}$.
Pople and coworkers reported the equations necessary for solving SCF
for DFT in the LCAO context up to generalized gradient approximation
(GGA) functionals in \citeref{Pople1992}; an analogous derivation was
also presented by Kobayashi \etal{} in \citeref{Kobayashi1991}.
The self-consistent implementation of meta-GGA functionals was later
described by Neumann, Nobes and Handy in \citeref{Neumann1996}.
Density functional calculations sometimes include also non-local
correlation contributions; self-consistent LCAO implementations
thereof have been reported by Vydrov and coworkers.\cite{Vydrov2008,
  Vydrov2009, Vydrov2010, Vydrov2010a}

Despite the progress in and widespread success of DFT, to our
knowledge, a uniform derivation of the SCF equations for HF and DFT
including all the necessary expressions for the elements of the
Kohn--Sham--Fock matrix up to the level of meta-GGA functionals has,
up to now, not been explicitly published in the literature.
This has likely contributed to the lack of complete support for
meta-GGA functionals in popular quantum chemistry programs; for
instance, \PsiFour{}\cite{Parrish2017} and \PySCF{}\cite{Sun2018} lack
support for meta-GGAs that depend on the Laplacian of the density such
as the Becke--Roussel exchange functional,\cite{Becke1989} for
example.
This paper, therefore, presents such a derivation, yielding
expressions of the DFT contributions to the Kohn--Sham--Fock matrix up
to the level of meta-GGA functionals in a consistent way, facilitating
the implementation of DFT in new programs.

The present derivation also has an obvious educational value.
Indeed, in what follows, HF and various flavors of DFT belonging to
different rungs of Jacob's Ladder\cite{Perdew2001}---the local spin
density approximation (LDA), the GGA and meta-GGA
approximations---will be explicitly described in a uniform notation,
making the similarities and dissimilarities between the approaches
crystal clear.
Facilitated by the uniform derivation, we will discuss key issues and
features in the HF and DFT methodologies that arise from the
mathematical formulation.

First, the basis set expansion of the molecular orbitals and the
electron density is written out in \secref{basisexp}.
Then, the energy expression for HF and DFT is presented in
\secref{energyexpr}, with a brief explanation of their physical
content.
The HF and DFT energy is shown to be invariant to rotations of the
occupied and of the virtual orbitals in \secref{unitinv}, allowing the
construction of localized orbitals.
The possibilities and drawbacks of spin-restricted calculations are
discussed in \secref{spinrestr}.
The finite-basis SCF equations are derived as generalized eigenvalue
equations in \secref{scfeqs}.
It is shown that the general eigenvalue equations can be reduced into
normal eigenvalue equations by a transformation to an orthonormal
basis in \secref{solscf}, and that linear dependencies in the basis
can be eliminated on the way.
The reason why the solution of the SCF equations amounts to a
minimization of the total energy is rationalized in
\secref{whyminimize}.
Direct minimization methods are briefly introduced and stability
analysis discussed in \secref{directmin}.
The SCF method and direct minimization are contrasted in
\secref{scfvsdirect}.
Finally, the contributions to the Kohn--Sham--Fock matrix arising from
various-rung DFT functionals are listed in \secref{dftcontr}.
The article concludes with a brief summary and discussion in \secref{summary}.
Atomic units are used throughout the text.

\section{Basis set expansion \label{sec:basisexp}}
In the HF and DFT approaches, the electronic wave function is written
as a Slater determinant, in which the electrons occupy a set of MOs
$\varphi(\vec{r})$.
The MOs are expanded in terms of normalized AOs $\chi(\vec{r})$, which
are typically not orthonormal to each other
\begin{equation}
\int \dr \ \chi_{\mu}(\vec{r})  \chi_{\mu}(\vec{r})   = 1 \nonumber
\hspace{3cm} 
\int \dr \ \chi_{\mu}(\vec{r})  \chi_{\nu}(\vec{r})   = S_{\mu\nu} \neq \delta_{\mu\nu} \nonumber
\end{equation}
where $\delta_{\mu \nu}$ is the Kronecker delta.
Greek letters, $\mu, \nu, \lambda, \sigma, \eta$, $\zeta$ and $\theta$
will be used to identify the expansion functions $\chi(\vec{r})$.
The $\alpha$ (spin-up) and $\beta$ (spin-down) MOs are expanded
separately in terms of the AOs as
\begin{eqnarray}
\varphi^{\alpha}_{i}(\vec{r}) = \sum^{M}_{\mu=1} C^{\alpha}_{\mu i} \chi_{\mu}(\vec{r}) \label{eq:psia} \\
\varphi^{\beta}_{i}(\vec{r}) = \sum^{M}_{\mu=1} C^{\beta}_{\mu i} \chi_{\mu}(\vec{r}) \label{eq:psib}
\end{eqnarray}
Both the $\alpha$ and $\beta$ MOs are orthonormal to themselves
\begin{equation}
\int \dr \ \varphi^{\alpha}_{i}(\vec{r}) \varphi^{\alpha}_{j}(\vec{r})   = \delta_{ij}
\hspace{1cm} \textnormal{and}  \hspace{1cm}
\int \dr \ \varphi^{\beta}_{i}(\vec{r}) \varphi^{\beta}_{j}(\vec{r})   = \delta_{ij} \nonumber
\end{equation}
However, the $\alpha$ orbitals are generally not orthonormal to the $\beta$ orbitals:
\begin{equation}
\int \dr \ \varphi^{\alpha}_{i}(\vec{r}) \varphi^{\beta}_{j}(\vec{r})   \neq \delta_{ij} \nonumber
\end{equation}
Roman letters, $i$, $j$ and $k$ will be used to identify the MOs
$\varphi$.
Both $\varphi$ and $\chi$, as well as the LCAO coefficients
$C_{\mu i}$, are typically chosen to be real.
For easier readability, from now on $\sum^{M}_{\mu=1}$ will be
simplified to $\sum_{\mu}$.

The electron density plays a pivotal role in quantum chemistry. In
line with chemistry literature, $\rho(\vec{r})$ will be used to denote
the electron density at the point $\vec{r}$ in contrast to the physics
notation $n(\vec{r})$ which is customary in the DFT literature.
The total electron density is formed from the $\alpha$ and $\beta$
densities, $\rho_\alpha$ and $\rho_\beta$, as $\rho(\vec{r}) =
\rho_\alpha(\vec{r}) +\rho_\beta(\vec{r})$.
The spin-$\sigma$ electron density can be evaluated as
\begin{equation}
  \rho_\sigma (\vec{r})
  = \sum^{N_{\sigma}}_{i} \left| \varphi^{\sigma}_{i}(\vec{r}) \right|^2
  = \sum^{N_{\sigma}}_{i} \sum_{\mu\nu} C^{\sigma}_{\mu i} C^{\sigma}_{\nu i} \chi_{\mu}(\vec{r}) \chi_{\nu}(\vec{r})
  = \sum_{\mu\nu} P^\sigma_{\mu\nu} \chi_{\mu}(\vec{r}) \chi_{\nu}(\vec{r}) \label{eq:sdens}
\end{equation}
in which $N_{\sigma}$ is the number spin-$\sigma$ electrons in the
system and where the density matrix has been defined as
\begin{equation}
  P^{\sigma}_{\mu\nu} =
  \sum_{i}^{N_{\sigma}} C^{\sigma}_{\mu i} C^{\sigma}_{\nu i} \label{eq:P}
\end{equation}
As is evident from the form of \eqref{P}, the density matrices are
symmetric, $P^\sigma_{\mu \nu} = P^\sigma_{\nu \mu}$.
As was already mentioned above, the total electron density is obtained
from the sum of the $\alpha$ and $\beta$ densities.
Correspondingly, a total density matrix is given by
\begin{equation}
  {\bf P} = {\bf P}^\alpha + {\bf P}^\beta \label{eq:Ptot}
\end{equation}
from which the total density can be evaluated using a relation
analogous to \eqref{sdens}.

\section{Energy expression \label{sec:energyexpr}}
The starting point for the derivation is the non-relativistic energy
expression \cite{Roothaan1951, Pople1954, Kohn1965, Pople1992},
\begin{equation}
\label{eq:eqe}
E = \sum_{\mu\nu} P_{\mu\nu} H_{\mu\nu}
  + \frac{1}{2} \sum_{\mu\nu\lambda\sigma} P_{\mu\nu} P_{\lambda\sigma} \ERI{\mu\nu}{\lambda\sigma}
  - \frac{a}{2}\sum_{\mu\nu\lambda\sigma}
(P^{\alpha}_{\mu\lambda} P^{\alpha}_{\nu\sigma}
+P^{\beta}_{\mu\lambda} P^{\beta}_{\nu\sigma}) \ERI{\mu\nu}{\lambda\sigma}
  + b \int{f(\vec{r}) \dr}
\end{equation}
where the electron repulsion integral \ERI{\mu\nu}{\lambda\sigma} is
defined as
\begin{equation}
\ERI{\mu\nu}{\lambda\sigma} = \int \dro \int \drt
\ \chi_{\mu}(\vec{r_1}) \chi_{\nu}(\vec{r_1})
     \frac{1}{r_{12}}
\ \chi_{\lambda}(\vec{r_2}) \chi_{\sigma}(\vec{r_2}) \label{eq:eri}
\end{equation}
and $a$ and $b$ are constants that define the fraction of HF
exchange and the weight of the density functional approximation,
respectively.
The choice $a=1$ and $b=0$ corresponds to HF, whereas $a=0$ and $b=1$
yields a ``pure'' density functional without exact exchange such as
the Perdew--Burke--Ernzerhof functional\cite{Perdew1996}.
The choice $a \neq 0$ and $b \neq 0$ is the most general one, which
corresponds to a hybrid functional\cite{Becke1993a} which are popular
in quantum chemistry; perhaps the most famous example being the
historical B3LYP functional.\cite{Stephens1994}

The first term in \eqref{eqe}, which will be referred to as
$E^{H}$, describes the kinetic energy of the electrons and the
Coulombic attraction of the $N$ nuclei in the system, with the matrix
elements
\begin{equation}
H_{\mu\nu} = \int dr \ \chi_{\mu}(\vec{r})
\left(-\frac{1}{2}\nabla^{2} + \sum_{N} \frac{Z_{N}}{r_{N}}\right)
\chi_{\nu}(\vec{r}) \label{eq:coreH}
\end{equation}
The one-electron operator in \eqref{coreH} is commonly known as the
core Hamiltonian, and the resulting $E^H$ is the dominating
contribution to the total energy.

However, the core Hamiltonian lacks electronic interactions.
These are described by the second and third terms in \eqref{eqe},
which describe the classical Coulomb and the quantum mechanical
``exchange'' energy, and are referred to as $E^{J}$ and $E^{K}$,
respectively.
The $E^{J}$ contribution to the total energy can be straightforwardly
derived from the expression for the Coulomb repulsion between the
electrons described by the electron density $\rho(\vec{r})$
\begin{equation}
E^{J} = \frac{1}{2} \int d\vec{r_1} \int d\vec{r_2}
          \ \rho(\vec{r_1}) \frac{1}{r_{12}} \rho(\vec{r_2}) \nonumber
\end{equation}
whereas the expression for the exchange energy contribution $E^{K}$
can be obtained, for instance, using Slater's rules for a HF wave
function ($a=1$).

The final term in \eqref{eqe}, referred to as $E^\text{XC}$, describes
the DFT exchange-correlation contributions which alike $E^J$ and $E^K$
arise from electronic interactions, and is commonly written as
\begin{equation}
  E^\text{XC} = \int \dr \ f(\vec{r}) = \int \dr \ \rho(\vec{r}) \epsilon^\text{XC}(\vec{r}) \nonumber
\end{equation}
where $\epsilon^\text{XC}$ is the exchange-correlation energy density.
In general, $f(\vec{r})$ is a function of the electron density, its
derivatives and the kinetic energy, depending on which rung of Jacob's
Ladder \cite{Perdew2001} is used to the describe the
exchange-correlation effects.

\section{Unitary invariance \label{sec:unitinv}}

The ${\bf P}^\alpha$ and ${\bf P}^\beta$ matrices turn out to be
invariant to rotations of the occupied orbitals among themselves.
Rotating the molecular orbitals $\varphi$ by a orthogonal matrix ${\bf
  U}$ defines a new set of orbitals
\begin{equation}
\varphi^{\alpha '}_{i} = \sum_k^{N_\alpha} \varphi^{\alpha}_{k} U_{ki} \nonumber
\end{equation}
the MO coefficients of which can be obtained as
\begin{equation}
C^{\alpha '}_{\mu i} = \sum_{k}^{N_\alpha} C^{\alpha}_{\mu k} U_{ki}
\nonumber
\end{equation}
This can also be written in matrix notation as
\begin{equation}
\label{eq:cpuc}
\mathbf{C^{\alpha '} = C^{\alpha} U}
\hspace{1cm} \textnormal{or}  \hspace{1cm}
\mathbf{C^{\alpha '} \trans{U} = C^{\alpha}}
\end{equation}
The invariance to rotations in the occupied-occupied block is easy to prove, as
\begin{equation}
P^{\alpha '}_{\mu\nu} = \sum_{i}^{N_\alpha} C^{\alpha '}_{\mu i} C^{\alpha '}_{\nu i}
               = \sum_{ikl}^{N_\alpha} U_{ik} C^{\alpha}_{\mu k} U_{il} C^{\alpha}_{\nu l}
               = \sum_{kl}^{N_\alpha} \delta_{kl} C^{\alpha}_{\mu k} C^{\alpha}_{\nu l}
               = \sum_{k}^{N_\alpha} C^{\alpha}_{\mu k} C^{\alpha}_{\nu k} = P^{\alpha}_{\mu\nu} \label{eq:Prot}
\end{equation}
where we have used the orthogonality of ${\bf U}$, $\trans{\bf U} {\bf U} = {\bf 1} = {\bf U} \trans{\bf U}$.

The invariance to rotations in the occupied-occupied block can be used
to fashion localized orbitals, for instance using an unitary
optimization procedure.\cite{Lehtola2013a}
Although localized orbitals are not strictly speaking
observables---due to which several localization criteria have been
suggested in the literature\cite{Foster1960, Edmiston1963, Pipek1989,
  Lehtola2014}---they have been shown to offer an effective way to
study chemical reactions with ab initio calculations.\cite{Knizia2015,
  Liu2018a, Klein2019}

In addition to the occupied orbitals, in general there are also a
number of unoccupied orbitals, which are commonly known as virtual
orbitals.
The number of virtual orbitals in any given calculation depends on the
size of the basis set: the bigger the basis is, the more virtual
orbitals there are.
Because the virtual orbitals do not enter into the density matrix, the
HF and DFT energy expression, \eqref{eqe}, is also invariant to
rotations in the virtual-virtual block.
However, as will be seen below, the energy can be changed by mixing
virtual orbitals into the occupied orbitals.\cite{Head-Gordon1988a}
This approach provides another way to optimize the orbitals directly
with \eg{} a gradient descent method.
An example of such an algorithm is the geometric direct minimization
method described in \citeref{VanVoorhis2002}.
The steps involved in gradient descent methods are similar to those in
the SCF method, and we refer the interested reader to the vast
literature on direct minimization methods that are too many to
comprehensively cite here.

\section{Spin-restriction vs unrestriction \label{sec:spinrestr}}

The molecular orbitals are obtained from the requirement that they
minimize the total energy according to \eqref{eqe}. However, one must
first choose the used formalism.
The general choice is to use different orbitals for the $\alpha$ and
$\beta$ electrons, in which case a spin-unrestricted approach is
obtained.
The unrestricted approach is often used even in systems in which there
are an equal number of alpha and beta electrons, $N_\alpha = N_\beta$.
Although the spin-restricted and unrestricted descriptions often
reproduce matching results for such systems near the equilibrium, only
the unrestricted formalism is able to break bonds in general.
The reason for this is that when molecules are stretched past the
Coulson--Fischer point,\cite{Coulson1949} the optimal orbitals
spontaneously break spin symmetry, which can only be described in the
unrestricted formalism.
At variance, in the spin-restricted case the electrons occupy a common
set of $N=N_{\alpha}=N_{\beta}=(N_\alpha + N_\beta)/2$ orbitals.
Because the limitation of the orbitals to be the same for both spins,
$C_{\mu i} = C^{\alpha}_{\mu i}=C^{\beta}_{\mu i}$, yields less
variational freedom, it affords computational savings.
The spin-restricted density matrices [\eqref{P}] reduce to
\begin{equation}
P^{\alpha}_{\mu\nu} = P^{\beta}_{\mu\nu} = \frac{1}{2} P_{\mu\nu} =
\sum_{i}^{N} C_{\mu i} C_{\nu i} \label{eq:prhf}
\end{equation}
meaning \eg{} that the $\alpha$ and $\beta$ exchange terms in
\eqref{eqe} coincide and can be simplified.

Spin-restriction is also possible in the case in which $N_\alpha \neq
N_\beta$.
In this case, a restricted open-shell method is obtained.
Restricted open-shell methods are more involved than the spin-restricted and spin-unrestricted methods discussed in the present work.
Restricted open-shell methods have been extensively discussed in
\citeref{Krebs1999} to which we refer for further discussion.

Having chosen to use either spin-restricted or spin-unrestricted
orbitals, one can proceed to minimization of the energy expression in
\eqref{eqe}.
The energy expression depends only on the $\alpha$ and $\beta$ density
matrices ${\bf P}^\alpha$ and ${\bf P}^\beta$ and their sum ${\bf P}$.
The density matrices, in turn, are determined by the lowest $N_\alpha$
and $N_\beta$ molecular orbitals [\eqref{P}].

\section{Self-consistent field equations \label{sec:scfeqs}}

Because the energy expression in \eqref{eqe} only depends on the
density matrices $\mathbf{P}^\alpha$ and $\mathbf{P}^\beta$, it is
expedient to use the chain rule to write \eg{}
\begin{equation}
\label{eq:ddc}
\frac{\partial \  }{\partial C^{\alpha}_{\theta k}} = \sum_{\eta\zeta}
\frac{\partial P^{\alpha}_{\eta\zeta}}{\partial C^{\alpha}_{\theta k}}
\frac{\partial \                     }{\partial P^{\alpha}_{\eta\zeta}}
\end{equation}
where the partial derivative of the density matrix element
$P^{\alpha}_{\eta\zeta}$ is
\begin{eqnarray}
  \label{eq:dPdc}
\frac{\partial P^{\alpha}_{\eta\zeta}}{\partial C_{\theta k}} &=&
\frac{\partial \ }{\partial C^{\alpha}_{\theta k}} \sum^{N_{\alpha}}_i C^{\alpha}_{\eta
i} C^{\alpha}_{\zeta i} \nonumber \\
&=& \sum^{N_{\alpha}}_i \delta_{\theta\eta} \delta_{ki} C^{\alpha}_{\zeta i} +
    \sum^{N_{\alpha}}_i C^{\alpha}_{\eta i} \delta_{\theta\zeta} \delta_{ki} \nonumber \\
\label{eq:derivpc}
&=& \delta_{\theta\eta} C^{\alpha}_{\zeta k} + C^{\alpha}_{\eta k}
\delta_{\theta\zeta}
\end{eqnarray}
Because the $\beta$ orbitals are formally independent from the
$\alpha$ orbitals (even though the same orbitals will be used in a
spin-restricted formalism), the $\alpha$ orbital derivative of the
total density matrix also coincides with \eqref{dPdc}.
Moreover, since the energy expression, \eqref{eqe}, is symmetric with
respect to the $\alpha$ and $\beta$ densities---it does not matter
which spin we choose to be ``up''---we only have to derive the
derivatives for one spin and the other will follow by the symmetry.

Because of the chain rule, \eqref{ddc}, all we need are the density
matrix derivatives of the energy expression.
The first term of \eqref{eqe} yields simply
\begin{equation}
\label{eq:parteh}
\frac{\partial E^{H} }{\partial P^{\alpha}_{\eta \zeta}}
= \frac{\partial }{\partial P^{\alpha}_{\eta \zeta}} \sum_{\mu\nu} P_{\mu\nu} H_{\mu\nu}
= H_{\eta \zeta}
\end{equation}
Next, taking the partial derivative with respect to
$P^{\alpha}_{\eta\zeta}$ of the Coulomb and exchange terms in
\eqref{eqe} results in
\begin{eqnarray}
\label{eq:partej}
\frac{\partial E^{J}}{\partial  P^{\alpha}_{\eta\zeta}} &=&
 \sum_{\mu\nu} (P^{\alpha}_{\mu\nu}+P^{\beta}_{\mu\nu}) \ERI{\mu\nu}{\eta\zeta}
 = \sum_{\mu\nu} P_{\mu\nu} \ERI{\mu\nu}{\eta\zeta} = J_{\eta \zeta}
\end{eqnarray}
where ${\bf J}$ is known as the Coulomb matrix, and
\begin{eqnarray}
\label{eq:partek}
\frac{\partial E^{K}}{\partial  P^{\alpha}_{\eta\zeta}} &=&
-a \sum_{\mu\nu} P^{\alpha}_{\mu\nu} \ERI{\mu\eta}{\nu\zeta} = -a K^{\alpha}_{\eta \zeta}
\end{eqnarray}
where ${\bf K}^\alpha$ is the spin-$\alpha$ exchange matrix,
respectively.
The Coulomb and exchange matrices can be used to rewrite the
energy expression in \eqref{eqe} as
\begin{equation}
  \label{eq:JKenergy}
E = \sum_{\mu\nu} P_{\mu\nu} H_{\mu\nu}
  + \frac{1}{2} \sum_{\mu\nu} P_{\mu\nu} J_{\mu \nu}
  - \frac{a}{2}\sum_{\mu\nu}
(P^{\alpha}_{\mu \nu} K^{\alpha}_{\mu\nu}
+P^{\beta}_{\mu\nu} K^{\beta}_{\mu\nu})
+ b \int{f(\vec{r}) \dr}
\end{equation}
Note that the exchange-correlation term does not undergo
simplifications, because as will be seen later, unlike the Coulomb and
exact exchange terms the exchange-correlation term is not quadratic in
the density matrix.
For the time being, we will denote the partial derivative of $E^\text{XC}$
with respect to $P^{\alpha}_{\eta\zeta}$ as
\begin{equation}
\label{eq:partexc}
\frac{\partial E^\text{XC}}{\partial P^{\alpha}_{\eta\zeta}} =
b K^{\text{XC};\alpha}_{\eta \zeta}
\end{equation}
as the full expressions for ${\bf K}^{\text{XC};\alpha}$ will be
presented in \secref{dftcontr}.
Now, collecting the partial derivatives in
\eqrangeref{parteh}{partexc} gives us the density matrix derivatives
of the energy expression as
\begin{equation}
\label{eq:fock}
\frac{\partial E}{\partial P^{\sigma}_{\eta\zeta}} = H_{\eta
  \zeta} + J_{\eta \zeta} -a K^\sigma_{\eta \zeta} +b
K^{\text{XC};\sigma}_{\eta \zeta} = F^{\sigma}_{\eta\zeta}
\end{equation}
where we have identified the Kohn--Sham--Fock matrices ${\bf
  F}^\sigma$, where $\sigma$ denotes $\alpha$ or $\beta$.
Because the density matrices defined by \eqref{P} are symmetric, also
the Fock matrices are symmetric, $F^\sigma_{\eta \zeta} =
F^\sigma_{\zeta \eta}$.
Note that since the Fock matrices only depends on the density
matrices, they are also invariant to occupied-occupied and
virtual-virtual rotations, $\mathbf{F^{\sigma'} = F^\sigma}$.

Na\"ively, one would obtain the orbital derivative of the full energy
expression in \eqref{eqe} with \eqref{ddc, dPdc, fock}, and set it to
zero to yield an equation for the unknown expansion coefficients ${\bf
  C}^\alpha$.
However, the molecular orbitals cannot be varied freely---one must
make sure that the orbitals stay orthonormal during the variation.
For instance, the orthonormality condition for the $\alpha$ electrons
is
\begin{equation}
\label{eq:condit}
\int \varphi^{\alpha}_i(\vec{r}) \varphi^{\alpha}_j(\vec{r})\\dr = \delta_{ij}
\end{equation}
The way to enforce these conditions is to use Lagrangian multipliers
$\varepsilon_{ij}$.
That is, instead of the bare energy expression $E$, we will optimize the
Lagrangian
\begin{equation}
  L = E - \sum_{ij} \varepsilon_{ij}^\alpha \left[ \int \dr \phi_i^\alpha (\vec{r}) \phi_j^\alpha (\vec{r}) - \delta_{ij} \right] - \sum_{ij} \varepsilon_{ij}^\beta \left[ \int \dr \phi_i^\beta (\vec{r}) \phi_j^\beta (\vec{r}) -\delta_{ij} \right] \label{eq:lagr}
\end{equation}
where the sums over $i$ and $j$ run over all orbitals; that is, both occupied
and virtual ones.
We can see from \eqref{lagr} that the matrices of Lagrangian
multipliers $\boldsymbol{\varepsilon}^\alpha$ and
$\boldsymbol{\varepsilon}^\beta$ can be chosen to be symmetric.
For instance, if $\boldsymbol{\varepsilon}^\alpha$ contained a
symmetric part $\boldsymbol{\varepsilon}^\alpha_s$ and an
antisymmetric part $\boldsymbol{\varepsilon}^\alpha_a$,
$\boldsymbol{\varepsilon}^\alpha = \boldsymbol{\varepsilon}^\alpha_s +
\boldsymbol{\varepsilon}^\alpha_a$, the contribution from the
antisymmetric part would vanish because it is multiplied with the
orbital overlap that is symmetric.

Next, we can calculate $\partial L / \partial C^\alpha_{\theta k}$,
where $\partial E / \partial C^\alpha_{\theta k} $ is given by
\eqref{ddc, dPdc, fock} and the derivative of the constraint term is
given by
\begin{eqnarray}
  \frac {\partial} {\partial C^{\alpha}_{\theta k}} \left[ \int \dr \phi_i^\alpha (\vec{r}) \phi_j^\alpha (\vec{r}) -\delta_{ij} \right]
   &=& \sum_{ij} \varepsilon_{ij}^\alpha \frac {\partial} {\partial C^{\alpha}_{\theta k}} \sum_{\eta\zeta} \left[ C^\alpha_{\eta i} S_{\eta\zeta} C^\alpha_{\zeta j}- \delta_{ij} \right] \nonumber \\
&=& \sum_{ij} \varepsilon^\alpha_{ij} \sum_{\eta\zeta}
\left[ \delta_{\eta\theta} \delta_{ki} C^{\alpha}_{\zeta j} +
        C^{\alpha}_{\eta i} \delta_{\zeta\theta} \delta_{kj}
\right] S_{\eta\zeta} \nonumber \\
\label{eq:dortha}
&=& \sum_{\eta} \sum_{i} \varepsilon^\alpha_{ki} C^{\alpha}_{\eta i} S_{\theta\eta}
    +\sum_{\eta}  \sum_{i} \varepsilon^\alpha_{ik} C^{\alpha}_{\eta i}  S_{\eta\theta}
\end{eqnarray}
where on the third line dummy summation indices have been renamed from
$j$ to $i$ and $\zeta$ to $\eta$.
The derivative can be evaluated as
\begin{eqnarray}
  \frac {\partial L} {\partial C^\alpha_{\theta k}} &=&
  \sum_{\eta\zeta} \frac {\partial E} {\partial P^\alpha_{\eta \zeta}} \frac {\partial P^\alpha_{\eta \zeta}} {\partial C^\alpha_{\theta k}} - \sum_{ij}
  \varepsilon_{ij}^\alpha \frac {\partial} {\partial C^\alpha_{\theta k}}
  \int \dr \phi_i^\alpha (\vec{r}) \phi_j^\alpha (\vec{r}) \nonumber \\
  &=& \sum_{\eta\zeta} F^\alpha_{\eta \zeta}
  \left[ \delta_{\theta \eta} C^\alpha_{\zeta k} + C^\alpha_{\eta k} \delta_{\theta \zeta} \right]
  -\sum_{\zeta} \sum_{i} \varepsilon_{ki}^\alpha C^{\alpha}_{\zeta i}
  S_{\theta\zeta} -\sum_{\eta} \sum_{i} \varepsilon_{ik}^\alpha
  C^{\alpha}_{\eta i} S_{\eta\theta} \nonumber \\
  \label{eq:dLdc}
  &=& 2 \sum_{\eta} F^\alpha_{\theta \eta} C^\alpha_{\eta k}
  - 2 \sum_{\eta} \sum_{i} \varepsilon_{ki}^\alpha S_{\theta\eta} C^{\alpha}_{\eta i}
\end{eqnarray}
because ${\bf F}$ and $\boldsymbol{\varepsilon}$ are symmetric, and
dummy summation indices can be renamed.

The optimal orbitals satisfy the stationary condition $\partial L /
\partial C^\alpha_{\theta k}=0$ from which
\begin{eqnarray}
  \label{eq:symmtheta}
\sum_{\eta} F^\alpha_{\theta \eta} C^\alpha_{\eta k} = \sum_{\eta}
\sum_{i} S_{\theta\eta} C^{\alpha}_{\eta i} \varepsilon_{ik}^\alpha
\end{eqnarray}
\Eqref{symmtheta} can thus be written in matrix form as
\begin{equation}
\label{eq:focka}
\mathbf{ F^{\alpha} C^{\alpha} = S C^{\alpha} E^{\alpha}}
\end{equation}
where ${\bf E}^\alpha$ is the (symmetric) matrix of Lagrangian multipliers.

Because ${\bf E}^\alpha$ is symmetric, it can be diagonalized and it
has real eigenvalues.
Let us now assume that ${\bf U}^\alpha$ is an orthogonal matrix that
diagonalizes ${\bf E}^\alpha$
\begin{equation}
\label{eq:udiag}
E_{ij}^\alpha \to  E_{ij}^{\alpha'} =\sum_{kl} U_{ki}^\alpha E_{kl}^\alpha U_{lj}^\alpha = \delta_{ij} \varepsilon_{i}^\alpha
\end{equation}
where $\boldsymbol{\varepsilon}^\alpha$ are the eigenvalues.
Rotating the orbital coefficients ${\bf C}^\alpha$ via the matrix
${\bf U}^\alpha$ via \eqref{cpuc}, ${\bf C}^\alpha = {\bf C}^{\alpha '} \trans{({\bf
  U}^{\alpha})}$, leads to
\begin{equation}
{\bf F}^{ \alpha'} {\bf C}^{\alpha '} \trans{({\bf U}^{\alpha})} = {\bf S} {\bf C}^{\alpha '} \trans{({\bf U}^{\alpha})} {\bf E}^{\alpha} \nonumber
\end{equation}
and multiplying both sides of this equation from the right by $\mathbf{U}^\alpha$
produces
\begin{equation}
  \label{eq:Fprime}
\mathbf{ F' C^{\alpha '} = S C^{\alpha '} E^{\alpha '}}
\end{equation}
where according to \eqref{udiag} ${\bf E}^{\alpha'} = \trans{({\bf
    U}^{\alpha})} {\bf E}^\alpha {\bf U}^\alpha$ is a diagonal matrix
with elements $\varepsilon^{\alpha}_{i}$.

\Eqref{Fprime} is almost what we want, but it still has a special
catch: the orbital rotation by ${\bf U}$ may mean that the
Kohn--Sham--Fock matrix is no longer the same as it was, ${\bf F}^{\sigma'} \neq {\bf F}^\sigma$.
However, if we choose the form of ${\bf U}$ such that the
occupied-virtual (ov) and virtual-occupied (vo) blocks vanish
\begin{equation}
  \boldsymbol{U}=\left(\begin{array}{cc}
    \boldsymbol{U}_{\text{oo}} & \boldsymbol{U}_{\text{ov}}\\
    \boldsymbol{U}_{\text{vo}} & \boldsymbol{U}_{\text{vv}}
  \end{array}\right)=\left(\begin{array}{cc}
    \boldsymbol{U}_{\text{oo}} & \boldsymbol{0}\\
    \boldsymbol{0} & \boldsymbol{U}_{\text{vv}}
    \end{array}\right) \label{eq:blockU}
\end{equation}
then ${\bf U}$ only rotates occupied orbitals with occupied orbitals
and virtual orbitals with virtual orbitals, meaning that the orbital
rotation does not change the density matrix given in \eqref{Prot}, and
the Fock matrix also stays the same, ${\bf F}' = {\bf F}$.
(Occupied-virtual rotations, discussed in more detail in
\secref{directmin}, are in fact here forbidden: the SCF equations were
derived with the assumption that the energy is stationary, but this
condition would instantly be violated by such rotations.)
In this case, we obtain the Berthier--Pople--Nesbet\cite{Pople1954,
  Berthier1954, Kohn1965, Pople1992} equations for the orbital
coefficients
\begin{equation}
  \label{eq:popnes}
  \begin{cases}
  {\bf F}^{\alpha} {\bf C}^{\alpha} = {\bf S} {\bf C}^{\alpha} {\bf E}^\alpha \\
        {\bf F}^{\beta} {\bf C}^{\beta} = {\bf S} {\bf C}^{\beta} {\bf E}^\beta
  \end{cases}
\end{equation}
where the primes have become unnecessary and have been omitted for
simplicity, and the elements of the Kohn--Sham--Fock matrix
$\mathbf{F^{\alpha}}$ and $\mathbf{F^{\beta}}$ are given by
\eqref{fock}.
In the spin-restricted case \cite{Roothaan1951, Pople1992} the
$\alpha$ and $\beta$ molecular orbitals coincide, leading to identical
density matrices ${\bf P}^\alpha$ and ${\bf P}^\beta$, and identical
Fock matrices ${\bf F}^\alpha$ and ${\bf F}^\beta$.
In this case, the SCF equations simplify to the Roothaan--Hall form
\begin{equation}
\label{eq:rh}
  \mathbf{ F C = S C E }
\end{equation}
which was already mentioned in the Introduction.

\section{Solution of self-consistent field equations \label{sec:solscf}}

The Roothaan--Hall and Berthier--Pople--Nesbet expressions take the
form of a generalized eigenvalue equation.
The conventional way to solve these equations is to again re-express
the (unknown) orbital coefficients in terms of a matrix ${\bf X}$ as
\begin{equation}
  {\bf C} = {\bf X} \tilde{\bf C}
  \label{eq:Xtrans}
\end{equation}
Inserting \eqref{Xtrans} into the Roothaan--Hall equation, \eqref{rh},
yields
\begin{equation}
  {\bf F} {\bf X} \tilde{\bf C} = {\bf S} {\bf X} \tilde{\bf C} {\bf E}
  \nonumber
\end{equation}
which can be multiplied from the left with $\trans{\bf X}$ to yield
\begin{equation}
  \trans{\bf X} {\bf F} {\bf X} \tilde{\bf C} = \trans{\bf X} {\bf S} {\bf X} \tilde{\bf C} {\bf E}
  \nonumber
\end{equation}
This means that the orbital transform of \eqref{Xtrans} yields a new
generalized eigenvalue equation
\begin{equation}
  \tilde{\bf F} \tilde{\bf C} = \tilde{\bf S} \tilde{\bf C} {\bf E}
  \label{eq:neweig}
\end{equation}
where $\tilde{\bf F} = \trans{\bf X} {\bf F} {\bf X}$ and
$\tilde{\bf S} = \trans{\bf X} {\bf S} {\bf X}$.
Now, if we choose ${\bf X}$ in such a way that $\tilde{\bf S} = {\bf
  1}$, \eqref{neweig} reduces to a normal eigenvalue equation
\begin{equation}
  \tilde{\bf F} \tilde{\bf C} = \tilde{\bf C} {\bf E}
  \label{eq:eigval}
\end{equation}
which can be solved with standard techniques.
Then, the wanted orbital coefficients ${\bf C}$ can be calculated from
$\tilde{\bf C}$ using \eqref{Xtrans}.

If the basis set is well-conditioned, the matrix ${\bf X}$ can be
chosen as
\begin{equation}
  {\bf X} = {\bf V} {\bf \Lambda}^{-1/2} \trans{\bf V}
  \label{eq:symmorth}
\end{equation}
where ${\bf V}$ and ${\bf \Lambda}$ are the eigenvectors and
eigenvalues of ${\bf S}$
\begin{equation}
  {\bf S} = {\bf V} {\bf \Lambda} \trans{\bf V}
  \label{eq:Sdec}
\end{equation}
This procedure is known as symmetric
orthogonalization.\cite{Lowdin1970}

However, if a large LCAO basis is used, the atomic orbital basis
functions centered on different atoms may generate significant linear
dependencies in the basis, making the basis set expansion ambiguous.
These linear dependencies can be removed with the ``canonical''
orthonormalization procedure,\cite{Lowdin1956} in which
\begin{equation}
  {\bf X} = {\bf V}' {\bf \Lambda}'^{-1/2}
  \label{eq:canorth}
\end{equation}
where only those eigenvectors ${\bf V}_i$ with large enough
eigenvalues $\lambda_i \geq \tau$ are included.
The threshold $\tau$ is typically of the order of $\tau = 10^{-6}
\dots 10^{-5}$, and its value may have a noticable effect on \eg{} the
absolute energies that result from a SCF calculation.
Relative energies, however, should be less sensitive.
If no eigenvalues fall under the threshold, the symmetric and
canonical orthogonalization approaches become equivalent for the
purposes of SCF calculations in the case of a well-conditioned basis
set: both yield an orthonormal basis that yields the same variational
ground state energy.

Unnormalized basis sets can also be handled easily by the
orthogonalization procedure.
Although in principle it is not necessary to normalize the individual
basis functions before obtaining an orthonormal basis by
\eqref{symmorth, canorth}, computer linear algebra packages may fail
to find the eigenvalues and eigenvectors in a reliable fashion if the
basis functions have pronouncedly different norms.
Moreover, missing normalization of the basis set affects the
eigenvalues, which has repercussions for canonical orthogonalization.
These issues can be circumvented by normalizing the overlap matrix
${\bf S} \to {\bf S}' = {\bf n} {\bf S} {\bf n}$ where $n_{ij} =
S_{ii}^{-1/2} \delta_{ij}$ before using \eqref{symmorth,
  canorth}.\cite{Lehtola2019a, Lehtola2019b}
The orthogonalizer for the unnormalized basis set is obtained as
${\bf X} \to {\bf n} {\bf X}$; it is easy to see that this satisfies
the necessary condition $\trans {\bf X} {\bf S} {\bf X} = {\bf 1}$
even though the symmetricity of ${\bf X}$ for the case of
\eqref{symmorth} will be lost.

Even if ${\bf S}$ has been properly normalized, the use of the
symmetric or canonical orthogonalization procedures still requires
that the diagonalization of ${\bf S}$ is numerically stable.
However, whenever a large number of diffuse functions are used in a
calculation or two nuclei are close together, ${\bf S}$ may become
ill-conditioned due to too many linear dependencies in the molecular
basis set.
In such cases it is possible to reduce the size of the basis set
without losing a significant amount of accuracy by an automatic
procedure, see \citerefs{Lehtola2019f} and \citeref{Lehtola2020a} for
details.

\section{Why does the self-consistent field method minimize the energy? \label{sec:whyminimize}}

The SCF equations, \eqref{popnes} or \eqref{rh}, offer a way to solve
for the molecular orbitals described by ${\bf C}$ from a
Kohn--Sham--Fock matrix ${\bf F}$ by finding its eigenvectors from
\eqref{eigval}.
However, the Kohn--Sham--Fock matrix depends on the density matrices,
which are built from the molecular orbitals ${\bf C}$ according to the
Aufbau principle.
In the SCF procedure, one tries to find a self-consistent solution:
${\bf C}$ yields a Kohn--Sham--Fock matrix ${\bf F}$, whose
eigenvectors are ${\bf C}$.
The procedure starts from an initial guess for the orbitals or the
density matrices, which have been recently reviewed and benchmarked in
\citeref{Lehtola2019} to which we refer for further details.

Why does this diagonalization procedure correspond to minimization of
the Hartree--Fock/Kohn--Sham energy?
For simplicity, let us examine the case of HF theory.
The energy expression, \eqref{JKenergy}, can be written in this case
($a=1$, $b=0$) as
\begin{equation}
  \label{eq:ehf}
E = \text{Tr } {\bf P H} + \frac 1 2 \text{Tr } {\bf P J} - \frac 1 2 \text{Tr } {\bf P^\alpha K^\alpha} - \frac 1 2 \text{Tr } {\bf P^\beta K^\beta}
\end{equation}
The Fock matrix elements, \eqref{fock}, are given by
\begin{eqnarray}
  {\bf F}^{\alpha} &=& {\bf H} + {\bf J} - {\bf K}^\alpha \label{eq:hfa} \\
  {\bf F}^{\beta} &=& {\bf H} + {\bf J} - {\bf K}^\beta \label{eq:hfb}
\end{eqnarray}
\Eqref{ehf} can be rewritten with \eqref{hfa, hfb} as
\begin{equation}
  \label{eq:mine}
E = \frac 1 2 \text{Tr } {\bf P^\alpha (H+F^\alpha)} + \frac 1 2 \text{Tr } {\bf P^\beta (H+F^\beta)}
\end{equation}
Expanding the density matrices using \eqref{P} we see that
\eqref{mine} can be written as
\begin{equation}
  \label{eq:emo}
E = \frac 1 2 \sum_{i}^{N_\alpha} (h^\alpha_{ii} + f^\alpha_{ii}) + \frac 1 2 \sum_{i}^{N_\beta} (h^\beta_{ii}+f^\beta_{ii})
\end{equation}
where the core Hamiltonian and Fock matrices have been written in the
molecular orbital basis, ${\bf h}^\sigma = \trans{({\bf C}^\sigma)}
{\bf H} {\bf C}^\sigma$ and ${\bf f}^\sigma = \trans{({\bf C}^\sigma)}
{\bf F}^\sigma {\bf C}^\sigma$.

If one were to start the calculation from the core guess, then in this
case $\sum_i h^\alpha_{ii}$ and $\sum_i h^\beta_{ii}$ are minimized.
However, as discussed in \citeref{Lehtola2019}, this is a horrible
choice as it completely disregards electronic repulsion effects,
meaning that the $\sum_i f^\alpha_{ii}$ and $\sum_i f^\beta_{ii}$
terms are far from optimal.
The Roothaan step---obtaining new molecular orbitals by
diagonalization of the Fock matrix---results in a minimization of the
$\sum_i f^\alpha_{ii}$ and $\sum_i f^\beta_{ii}$ terms, as after
diagonalization only the lowest orbitals become populated and the sum
thus runs only over the lowest eigenvalues $f^\sigma_{ii}$.
After the update, $\sum_i h^\alpha_{ii}$ and $\sum_i h^\beta_{ii}$ no
longer yield their lowest possible values.
However, the increase in the value of $\sum_i h^\alpha_{ii} + \sum_i
h^\beta_{ii}$ should be much smaller than the decrease in the value of
$\sum_i f^\alpha_{ii} + \sum_i f^\beta_{ii}$, as the Fock matrices
${\bf f}^\alpha$ and ${\bf f}^\beta$ also contain the core
Hamiltonian.
It is thus seen that Roothaan's self-consistent field method, that is,
the iterative diagonalization of the Fock matrix minimizes the energy.

However, the minimization is only valid for a fixed potential ${\bf
  f}^\sigma$ in which the electrons are moving.
When the orbitals are changed---as happens when ${\bf f}$ is made
diagonal and its lowest eigenvectors occupied---a new Fock matrix
${\bf F}$ must be built and a new ${\bf f}$ constructed: the potential
also changes with the electron density.
If the orbitals were far from their optimal values, ${\bf P}$ and
therefore ${\bf F}$ may change quite radically by the orbital update.
This means that even though ${\bf f}$ was made diagonal in the
previous iteration, it is no longer diagonally dominant after it has
been updated.
Indeed, the straightforward iterative diagonalization procedure often
fails to converge for all but the simplest systems, because the
density tends to undergo large oscillations in the self-consistency
cycle.
Instead, the convergence of the fixed-point problem of finding a ${\bf
  C}$ that generates ${\bf F}$ that generates ${\bf C}$ must be
stabilized or accelerated in some way.
This can be achieved \eg{} by damping,\cite{Karlstrom1979, Cances2000}
level shifts,\cite{Saunders1973, Mitin1988, Domotor1989}, or
extrapolation.\cite{Pulay1980, Pulay1982, Kudin2002, Hu2010}
Fractional occupations can also be used in the initial iterations to
aid convergence.\cite{Rabuck1999}
The argument for why density functional calculations converge
similarly to HF with the iterative Roothaan procedure is somewhat less
obvious, because unlike HF the exchange-correlation functional is not
generally quadratic in the density.
However, the total energy expression \emph{is} quadratic also in DFT
sufficiently close to an extremal point, as is easily seen by a Taylor
expansion of \eqref{eqe}.
In practice the iterative procedure works well also for DFT, whose
contributions to the Kohn--Sham--Fock matrix we will discuss in the
last section.

\section{Direct minimization of the energy \label{sec:directmin}}

Instead of solving the orbitals from the SCF equations, which were
obtained above from the stationary condition for the energy under the
constraint of orthonormal orbitals, the orbitals can also be optimized
by a direct minimization of the energy.
As was discussed above, the energy expression of \eqref{eqe} is
invariant to occupied-occupied and virtual-virtual rotations.
This means that if we have $o$ occupied orbitals and $v$ virtual
orbitals from some initial guess (see alternatives in
\citeref{Lehtola2019}), we can consider the energy to be a function of
a set of $ov$ rotation angles\cite{Head-Gordon1988a} by examining a
rotation of the orbitals via \eqref{cpuc} by an orthogonal matrix
\begin{equation}
{\bf U}^\sigma(\boldsymbol{\theta})=\exp\left(\begin{array}{cc} \boldsymbol{0} &
  \boldsymbol{\theta}^\sigma\\ -\trans{(\boldsymbol{\theta}^\sigma)} &
  \boldsymbol{0}
\end{array}\right) \label{eq:ovrot}
\end{equation}
where $\boldsymbol{\theta}$ is an $o \times v$ matrix containing the
rotation angles.
The rotation matrix determined by \eqref{ovrot} reduces to an identity
matrix for vanishing rotation parameters, $\boldsymbol{\theta}={\bf
  0}$.
Because the rotation matrix of \eqref{ovrot} is orthogonal, it
automatically preserves the orthonormality of the orbitals, and
special tricks \ie{} Lagrangian multipliers are not needed to enforce
this behavior.

The change in the density matrix is given by
\begin{equation}
  \frac {\partial P^\sigma_{\eta \zeta}} {\partial \theta^\sigma_{ia}}
  = \sum_{k}^{N_{\sigma}} \left[ \frac {\partial C^{\sigma}_{\eta k}}
    {\partial \theta^\sigma_{ia}} C^\sigma_{\zeta k} + C^\sigma_{\eta k}
    \frac {\partial C^{\sigma}_{\zeta k}} {\partial \theta^\sigma_{ia}} \right]
\end{equation}
How do the orbital coefficients change?
Remembering that the first $N_\sigma$ orbitals are occupied, and the
rest are virtual, we can write
$\boldsymbol{C}^{\sigma}=(\begin{array}{cc}
  \boldsymbol{C}_{o}^{\sigma} &
  \boldsymbol{C}_{v}^{\sigma})\end{array}$.
After an infinitesimal rotation $\boldsymbol{\theta}\approx {\bf 0}$,
the occupied orbitals change into $\boldsymbol{C}_o^{\sigma'} \approx
\boldsymbol{C}_o^{\sigma} - \boldsymbol{C}_v^{\sigma}
\trans{(\boldsymbol{\theta}^\sigma)}$, that is
\begin{equation}
  C^{\sigma'}_{\eta k} = C^{\sigma}_{\eta k} - \sum_{b\text{ virtual}}
  C^{\sigma}_{\eta b} \theta^\sigma_{kb}
\end{equation}
from which ${\partial C^{\sigma}_{\eta k}} / {\partial
  \theta^\sigma_{ia}} = -C^\sigma_{\eta a} \delta_{ik}$.
Now the gradient of the energy with respect to rotation of the current
set of orbitals can be obtained as
\begin{equation}
  \frac {\partial E} {\partial \theta_{ia}^\sigma}
  = \sum_{\eta\zeta} \frac {\partial E} {\partial P^\sigma_{\eta
      \zeta}} \frac {\partial P^\sigma_{\eta \zeta}} {\partial
    \theta^\sigma_{ia}}
  = \sum_{\eta\zeta} F^\sigma_{\eta \zeta} \sum_k^{N_\sigma} \left[ \frac {\partial C^{\sigma}_{\eta k}}
    {\partial \theta^\sigma_{ia}} C^\sigma_{\zeta k} + C^\sigma_{\eta k}
    \frac {\partial C^{\sigma}_{\zeta k}} {\partial \theta^\sigma_{ia}} \right]
  = - \sum_{\eta\zeta} F^\sigma_{\eta \zeta} \left[ C^\sigma_{\eta a}  C^\sigma_{\zeta i} + C^\sigma_{\eta i}
    C^\sigma_{\eta a} \right] = -2 f^\sigma_{ia} \label{eq:dEdtheta}
\end{equation}
where ${\bf f}^\sigma = \trans{(\bf{C}^\sigma)} {\bf F}^\sigma {\bf
  C}^\sigma$ is the Fock matrix in the MO basis.
Direct minimization of \eqref{eqe} can then be pursued using
\eqref{dEdtheta} with \eg{} gradient descent methods.
However, a proper preconditioning of the search direction is essential
in order for the algorithm to be usable; see \eg{} the geometric
direct minimization method described in \citeref{VanVoorhis2002}.

\section{SCF vs direct minimization\label{sec:scfvsdirect}}

Having described two alternative ways for solving the orbitals, we can
discuss their advantages and disadvantages.
The self-consistent field method is hard to beat for systems where
convergence is straightforward: a suitably stabilized and accelerated
SCF procedure often converges within a 10 to 20 iterations when a
suitable\cite{Lehtola2019} initial guess has been provided.
However, when the gap between the highest occupied and lowest
unoccupied orbital is small, which commonly occurs in \eg{} first-row
transition metal complexes, the SCF procedure may become extremely
slow, oscillate between two or more solutions, converge to a
higher-lying solution, or a saddle-point solution.
Namely, it is critically important to realize that even if the orbital
gradient vanishes, or equivalently, that the SCF equations are
fulfilled, this does not mean that the energy expression \eqref{eqe}
truly has been minimized.
Because there are typically several occupied as well as virtual
orbitals, the minimization problem involves a large number of degrees
of freedom.
In multivariate calculus, a vanishing gradient only means that the
orbitals correspond to some kind of extremum of the energy: a local
minimum, a saddle-point solution, or even a local maximum, although
the lattermost is highly improbable in SCF calculations.

In contrast to the sometimes erratic behavior of the SCF method,
direct minimization based on orbital rotations is guaranteed to
converge onto an extremal point $f_{ia}=0$ per the theory of numerical
analysis; this is of great worth when studying systems with
complicated electronic structures for which conventional SCF
algorithms fail.
However, more predictable convergence does not come for free: the
downside of direct minimization methods is that they carry a higher
computational cost due to \eg{} the use of line searches in the
orbital optimization.
Direct minimization methods can also be formulated at the second
order, yielding more robust convergence to a local minimum at the cost
of more computational resources per iteration.\cite{Douady1979,
  Douady1980, Head-Gordon1989}
Because direct minimization methods are based on an explicit rotation
of the orbitals, they are able to always follow the same solution at
variance to SCF methods where the orbital occupations are typically
reset at every iteration according to the Aufbau principle.
Because of this, direct minimization can lead to a solution where the
Aufbau rule is violated, that is, the highest occupied orbital lies
higher in energy than the lowest unoccupied orbital.
Direct minimization methods can also be straightforwardly applied even
in more complicated electronic structure theories than self-consistent
field theory.
Such methods include explicit dependence on the molecular orbitals, as
discussed by one of the present authors in \citerefs{Lehtola2014a,
  Lehtola2015a, Lehtola2016} for the Perdew--Zunger self-interaction
correction\cite{Perdew1981} which depends explicitly on the $N_\sigma$
occupied orbitals, and \citeref{Lehtola2017} for the perfect
quadruples\cite{Parkhill2009} and perfect hextuples\cite{Parkhill2010}
models that also depend on the virtual orbitals.

In order to check the character of the extremum found by the SCF
procedure or a direct minimization method, it is necessary to continue
the analysis to second-order changes in the energy with respect to the
orbital rotations by finding the lowest eigenvalue of the Hessian
matrix: if it is negative, rotating the orbitals in the direction of
the corresponding eigenvector will result in a further decrease of the
energy.
Whenever post-HF calculations are performed, or benchmark-quality
values are sought at the SCF level, stability
analysis\cite{Seeger1977, Bauernschmitt1996} should be used to
guarantee that the wave function indeed corresponds to a local
minimum.
Alternatively, trust-region methods\cite{Francisco2004, Thogersen2004,
  Thogersen2005} can be employed to ensure that the orbitals converge
onto a true local minimum.

As always in the minimization of multivariate functions, locating the
global minimum is difficult, and typically the best one can hope for
is to find a local minimum.
Some systems permit several local electronic minima: for instance,
charge transfer complexes may allow both a neutral \ce{X\bond{...}Y}
as well as an ionic \ce{X+\bond{...}Y-} solution.
Finding such physically motivated solutions is often straightforward
by suitable manipulations of the initial guess, for instance, by
constructing guesses via the superposition of atomic
potentials\cite{Almlof1982, VanLenthe2006} with the correct atomic
charges.
Sometimes it may also be interesting to locate saddle-point solutions,
which have physical interpretations as excited states.
Specific excited states can be explored within the SCF approach by
replacing Aufbau population of the orbitals with overlap
criteria,\cite{Gilbert2008} or by direct minimization by replacing the
energy with the square of the gradient;\cite{Hait2020} for instance,
such an approach has been recently shown to predict highly accurate
core spectra.\cite{Hait2020a}
The full space of SCF solutions can be explored via \eg{}
meta-dynamics.\cite{Thom2008}

\section{Density functional contributions to Kohn--Sham--Fock matrix \label{sec:dftcontr}}

In \secref{scfeqs} we derived expressions for the Kohn--Sham--Fock
matrix elements for all but the density functional contribution
\begin{equation}
  E^\text{XC} = b \int{f(\vec{r}) \dr}
\end{equation}
which we will consider next.
Hundreds of density functionals $f(\vec{r})$ of various forms have
been published in the literature in the recent
decades,\cite{Mardirossian2017a} 
and offering a comprehensive selection thereof poses a considerable
challenge to quantum chemistry software developers.
This problem is further exacerbated by the need to keep track with the
several new functionals still being published every year.
Moreover, the density functionals $f(\vec{r})$ typically carry
extremely complicated functional forms, making their correct
implementation painstaking work.
The implementation is made even more difficult by the need to compute
the first derivatives of $f(\vec{r})$ for the SCF procedure, as well
as several higher-order ones for \eg{} the calculation of various
properties.

Fortunately, these challenges have been obviated by freely available,
portable standard implementations such as \LibXC{}\cite{Lehtola2018}
and \XCFun{}\cite{Ekstrom2010}.
The \LibXC{} software package strives to implement \emph{all DFT
  functionals published in the literature}, and provides a uniform
interface to functionals of various forms.
At present, \LibXC{} is used by 30 electronic structure programs based
on various numerical approaches that range from basis set approaches
(Gaussian-type orbitals, Slater-type orbitals, numerical atomic
orbitals, finite elements, plane waves) to finite difference
procedures.
New functionals only have to be added once to \LibXC{}, meaning the
library is easily kept up to date, after which they become available
to all programs that support the corresponding rung of functionals on
Jacob's ladder\cite{Perdew2001}.
Next, we will derive the equations necessary to implement the various
rungs in the variational basis set approach.

\subsection{LDA functionals \label{sec:lda}}
The simplest density functional approximations (DFAs), belonging to
the first rung of Jacob's Ladder\cite{Perdew2001}, are generally
referred to as local (spin) density approximations (LDAs).
These are functions of only the electron density\cite{Kohn1965}
\begin{equation}
f(\vec{r}) = f(\rho_\alpha(\vec{r}),\rho_\beta(\vec{r})) \label{eq:lda}
\end{equation}
such as the LDA exchange functional\cite{Bloch1929, Dirac1930}
\begin{equation}
f(\vec{r}) = -\frac{3}{2} \left( \frac{3}{4\pi} \right)^{\frac{1}{3}}
          \left( \rho_{\alpha}^{4/3}(\vec{r})+\rho_{\beta}^{4/3}(\vec{r}) \right) \nonumber
\end{equation}
Assuming $f$ has the form of \eqref{lda}, the resulting contribution
to the Kohn--Sham--Fock matrix $F^{\alpha;\text{XC}}_{\mu\nu} =
\partial E^\text{XC} / \partial P^\alpha_{\mu \nu}$ can be evaluated
using \eqref{P} for the densities at point $\vec{r}$
\begin{equation}
\rho_{\sigma}(\vec{r}) = \sum_{\mu\nu} P^{\sigma}_{\mu\nu}
\chi_{\mu}(\vec{r}) \chi_{\nu}(\vec{r}) \label{eq:rho}
\end{equation}
as\cite{Pople1992}
\begin{equation}
  F^{\alpha;\text{LDA}}_{\mu\nu} = \frac{\partial E^\text{XC}}{\partial
    P^{\alpha}_{\mu\nu}} = b \int\dr \ \frac{\partial f(\vec{r})
  }{\partial \rho_{\alpha}} \frac{\partial \rho_{\alpha}}{\partial
    P^{\alpha}_{\mu\nu}}
  = b \int\dr \ \frac{\partial f(\vec{r})}{\partial \rho_{\alpha}}
  \chi_{\mu}(\vec{r}) \chi_{\nu}(\vec{r})
\label{eq:frung1}
\end{equation}
with ${\bf F}^{\beta;\text{LDA}}$ having an analogous expression.
Note that if the integral is evaluated using numerical quadrature,
\begin{equation}
  F^{\alpha;\text{LDA}}_{\mu\nu} \approx b \sum_i w_i \ \frac{\partial
    f(\vec{r}_i)}{\partial \rho_{\alpha}} \chi_{\mu}(\vec{r_i})
  \chi_{\nu}(\vec{r_i})
\label{eq:numquad}
\end{equation}
Becke's multigrid approach\cite{Becke1988a} and further developments
thereof being the standard approach in LCAO programs, the expression
of \eqref{numquad} can be most efficiently formulated with matrix
products.
Storing the values of the basis functions at the quadrature points as
a matrix
\begin{equation}
  X_{\mu i}=\chi_\mu(\vec{r}_i)
  \nonumber
\end{equation}
and defining a scaled version thereof as
\begin{equation}
  Y^\alpha_{\mu i}= w_i \frac{\partial f(\vec{r}_i)}{\partial \rho_{\alpha}}
  \chi_\mu(\vec{r}_i) \nonumber
\end{equation}
the Fock matrix contribution can be evaluated as simply as
\begin{equation}
  {\bf F}^{\alpha;\text{LDA}} = b {\bf X} \trans{({\bf Y}^\alpha)}
  \nonumber
\end{equation}
which is orders of magnitude faster than a simple \emph{for} loop
based algorithm.

\subsection{GGA functionals \label{sec:gga}}
The second rung of Jacob's Ladder \cite{Perdew2001} is referred to as
the Generalised Gradient Approximation\cite{Perdew1986a} (GGA).
Density functional approximations on this rung also depend on the
derivatives of the density
\begin{equation}
     f(\vec{r}) = f(\rho_{\alpha} ,\rho_{\beta}
                    ,\gamma_{\alpha\alpha},\gamma_{\alpha\beta},\gamma_{\beta\beta}
                   ) \nonumber
\end{equation}
via the reduced gradients $\gamma_{\alpha\alpha}=\nabla \rho_{\alpha}
\cdot \nabla \rho_{\alpha}$, $\gamma_{\alpha\beta}=\nabla
\rho_{\alpha} \cdot \nabla \rho_{\beta}$, and
$\gamma_{\beta\beta}=\nabla \rho_{\beta} \cdot \nabla \rho_{\beta}$,
with the gradient of the density $\nabla \rho_{\sigma}$ being
determined by
\begin{equation}
\nabla \rho_{\sigma}
= \sum_{\mu\nu} P^{\sigma}_{\mu\nu} \nabla[\chi_{\mu}(\vec{r})\chi_{\nu}(\vec{r})]
= \sum_{\mu\nu} P^{\sigma}_{\mu\nu}
    [\chi_{\nu}(\vec{r}) \nabla \chi_{\mu}(\vec{r})
    +\chi_{\mu}(\vec{r}) \nabla \chi_{\nu}(\vec{r})
    ] \nonumber
\end{equation}
The GGA contribution to the Fock matrix is given by\cite{Pople1992}
\begin{eqnarray}
F^{\alpha;\text{GGA}}_{\mu\nu} &=& \frac{\partial E^\text{XC}}{\partial P^{\alpha}_{\mu\nu}} \nonumber \\
&=&
\int\dr\  \left[ \frac{\partial f(\vec{r})   }{\partial \rho_{\alpha}}
                        \frac{\partial \rho_{\alpha}}{\partial P^{\alpha}_{\mu\nu}}
                      + \frac{\partial f(\vec{r})           }{\partial \gamma_{\alpha\alpha}}
                        \frac{\partial \gamma_{\alpha\alpha}}{\partial \nabla \rho_{\alpha} }
                        \cdot \frac{\partial \nabla \rho_{\alpha  }}{\partial P^{\alpha}_{\mu\nu} }
                      + \frac{\partial f(\vec{r})           }{\partial \gamma_{\alpha\beta} }
                        \frac{\partial \gamma_{\alpha\beta }}{\partial \nabla \rho_{\alpha} }
                        \cdot \frac{\partial \nabla \rho_{\alpha }}{\partial P^{\alpha}_{\mu\nu}  }
                \right] \nonumber \\
&=&
F^{\alpha;\text{LDA}}_{\mu\nu} + \int\dr\  \left( 2 \frac{\partial f(\vec{r}) }{\partial \gamma_{\alpha\alpha}} \nabla \rho_{\alpha}(\vec{r})
               + \frac{\partial f(\vec{r}) }{\partial \gamma_{\alpha\beta }} \nabla \rho_{\beta}(\vec{r})
       \right) \cdot
    [\chi_{\nu}(\vec{r}) \nabla \chi_{\mu}(\vec{r})
    +\chi_{\mu}(\vec{r}) \nabla \chi_{\nu}(\vec{r})
    ]
\label{eq:frung2}
\end{eqnarray}
The $\beta$ expression can be obtained by switching $\alpha$ and
$\beta$ in \eqref{frung2}.
In the restricted case, $E^\text{XC} = \int\dr f(\rho,\gamma)$ with
$\gamma=\nabla \rho \cdot \nabla \rho$, which leads to the DFT
contributions to ${\bf F}^{\text{GGA}}$ given by the simpler expression
\begin{equation}
\label{eq:frung2r}
F^\text{GGA}_{\mu\nu} =
\int\dr\  \left[ \frac{\partial f(\vec{r})   }{\partial \rho}
                             \chi_{\mu}(\vec{r}) \chi_{\nu}(\vec{r})
     + 2 \frac{\partial f(\vec{r}) }{\partial \gamma} \nabla \rho(\vec{r})
       \cdot  \left[ \chi_{\nu}(\vec{r}) \nabla \chi_{\mu}(\vec{r})
          +\chi_{\mu}(\vec{r}) \nabla \chi_{\nu}(\vec{r})
          \right] \right]
\end{equation}
Practical implementations of \eqref{frung2, frung2r} can again be
formulated using matrix products.

\subsection{Meta-GGA functionals \label{sec:mgga}}
On the third rung on Jacob's Ladder\cite{Perdew2001} are the meta-GGA
(mGGA) approximations
\begin{equation}
     f(\vec{r}) = f(\rho_{\alpha} ,\rho_{\beta}
                    ,\gamma_{\alpha\alpha},\gamma_{\alpha\beta},\gamma_{\beta\beta}
                    ,\tau_{\alpha},\tau_{\beta}
                    ,\nabla^{2} \rho_{\alpha},\nabla^{2} \rho_{\beta}
                   ) \nonumber
\end{equation}
in which $\tau_{\sigma}$ and $\nabla^{2} \rho_{\sigma}$ are obtained as
\begin{eqnarray}
\tau_{\sigma} &=& \frac{1}{2} \sum^{n_{\sigma}}_{i} |\nabla \varphi_{i}(\vec{r})|^{2}
               =  \frac{1}{2} \sum^{n_{\sigma}}_{i}
                       \nabla \varphi_{i}(\vec{r}) \cdot  \nabla \varphi_{i}(\vec{r})
               =  \frac{1}{2} \sum^{n_{\sigma}}_{i}
                              \sum_{\mu\nu} C^{\sigma}_{\mu i} C^{\sigma}_{\nu i}
                              \nabla \chi_{\mu}(\vec{r}) \cdot \nabla \chi_{\nu}(\vec{r}) \nonumber \\
              &=& \frac{1}{2} \sum_{\mu\nu} P^{\sigma}_{\mu\nu}
                              \nabla \chi_{\mu}(\vec{r}) \cdot  \nabla \chi_{\nu}(\vec{r}) \label{eq:tau} \\
\nabla^{2} \rho_{\sigma} &=& \sum_{\mu\nu} P^{\sigma}_{\mu\nu}
           \nabla^{2} [\chi_{\mu}(\vec{r}) \chi_{\nu}(\vec{r})]    \nonumber \\
                         &=& \sum_{\mu\nu} P^{\sigma}_{\mu\nu}
           [  \chi_{\mu}(\vec{r}) \nabla^{2} \chi_{\nu}(\vec{r})
           +2 \nabla \chi_{\mu}(\vec{r}) \cdot \nabla \chi_{\nu}(\vec{r})
           +  \chi_{\nu}(\vec{r}) \nabla^{2} \chi_{\mu}(\vec{r})
           ]                                                       \label{eq:lapl} \\
                         &=& \sum_{\mu\nu} P^{\sigma}_{\mu\nu}
           [  \chi_{\mu}(\vec{r}) \nabla^{2} \chi_{\nu}(\vec{r})
           +  \chi_{\nu}(\vec{r}) \nabla^{2} \chi_{\mu}(\vec{r})
           ] + 4 \tau_\sigma                                                      \nonumber
\end{eqnarray}
The meta-GGA contributions to the Kohn--Sham--Fock matrix are
straightforwardly obtained as\cite{Neumann1996}
\begin{eqnarray}
F^{\alpha;\text{mGGA}}_{\mu\nu} &=& \frac{\partial E^\text{XC}}{\partial P^{\alpha}_{\mu\nu}} \nonumber \\
                          &=& F^{\alpha;\text{GGA}}_{\mu\nu} + \int\dr\ \left[\
           \frac{\partial f(\vec{r})  }{\partial \tau_{\alpha}      }
           \frac{\partial \tau_{\alpha}}{\partial P^{\alpha}_{\mu\nu}}
          +
           \frac{\partial f(\vec{r})                       }{\partial \nabla^{2} \rho_{\alpha}(\vec{r})}
           \frac{\partial \nabla^{2} \rho_{\alpha}(\vec{r})}{\partial P^{\alpha}_{\mu\nu}}
                                          \ \right] \nonumber \\
\label{eq:frung3}
                          &=& F^{\alpha;\text{GGA}}_{\mu\nu} + \int\dr\ \left[\
  \left( \frac{1}{2} \frac{\partial f(\vec{r})}{\partial \tau_{\alpha}}
  + 2 \frac{\partial f(\vec{r})}{\partial \nabla^{2} \rho_{\alpha}(\vec{r})} \right)  
     \nabla \chi_{\mu}(\vec{r}) \cdot \nabla \chi_{\nu}(\vec{r}) \right. \nonumber \\
     && \ \ \ \ \ \ \ \ \ \ \ \ \ \ \ \ \ \ \ \ \      +
                       \left. \frac{\partial f(\vec{r})                       }{\partial \nabla^{2} \rho_{\alpha}(\vec{r})}
           [ \chi_{\mu}(\vec{r}) \nabla^{2} \chi_{\nu}(\vec{r})
           + \chi_{\nu}(\vec{r}) \nabla^{2} \chi_{\mu}(\vec{r})
           ]
                                             \right]
\end{eqnarray}
which can again be expressed in terms of matrix products to achieve
faster quadrature.
In the restricted case, the expressions remain formally the same but
the quantities correspond to the total electron density.

\subsection{Range-separated hybrid functionals \label{sec:rangesep}}
As was mentioned before, the use of non-zero values for the constants
$a$ and $b$ in equation (\ref{eq:eqe}) allows the inclusion of exact
exchange effects in a DFT calculation.
These functionals represent the fourth rung of Jacob's Ladder
\cite{Perdew2001}, and are generally referred to as hybrid DFT
calculations.
A further development on hybrid functionals are range-separated
hybrids,\cite{Savin1995, Leininger1997} in which the interelectronic
interaction is divided into a short-range (sr) and a long-range (lr)
part with a resolution of the identity
\begin{equation}
  \label{eq:rangesep}
  \frac{1}{r_{12}} = \frac{\phi^\text{sr}(r_{12})}{r_{12}}
  + \frac{\phi^\text{lr}(r_{12})}{r_{12}}
\end{equation}
where $\phi^\text{sr}(r_{12})+\phi^\text{lr}(r_{12})=1$.
The rationale for range separation is that since density functional
approximations for the exchange are based only on local information
about the density, they fail to reproduce accurate estimates for
charge transfer processes, for example.
Separating the interaction by range per \eqref{rangesep} leads to a
hybrid exchange functional that has four contributions
\begin{equation}
  \label{eq:rsx}
  E^\text{X} = a^\text{sr} E^{\text{sr-HF}} + a^\text{lr}
  E^{\text{lr-HF}} + b^\text{sr} E^{\text{sr-DFT}} + b^\text{lr}
  E^{\text{lr-DFT}} = a^\text{sr} E^{\text{sr-HF}} + a^\text{lr}
  E^{\text{lr-HF}} + E^{\text{DFT}}
\end{equation}
where we have stressed that since the DFT contributions are evaluated
based only on the density (and possibly its derivatives), $b^\text{sr}
E^{\text{sr-DFT}} + b^\text{lr} E^{\text{lr-DFT}}$ is nothing but a
definition of a new density functional.
In contrast, the HF contributions to the energy and the
Kohn--Sham--Fock matrix have to be evaluated separately with
range-separated ERIs
\begin{equation}
  \ERI{\mu\nu}{\lambda\sigma}^\text{sr/lr} = \int \dro \int \drt
\ \chi_{\mu}(\vec{r_1}) \chi_{\nu}(\vec{r_1})
     \frac{\phi^\text{sr/lr}(r_{12})}{r_{12}}
\ \chi_{\lambda}(\vec{r_2}) \chi_{\sigma}(\vec{r_2}) \label{eq:rseri}
\end{equation}
Several kinds of range-separation kernels $\phi^\text{sr/lr}(r)$ have
been proposed; however, the error function based kernel
$\phi^\text{sr}(r;\omega)=\text{erfc }(\omega r)$,
$\phi^\text{lr}(r;\omega)=\text{erf }(\omega r)$, where $\omega$ is
the range-separation parameter, is by far the most commonly used one
because it is exceedingly simple to implement in codes employing a
plane-wave or Gaussian basis set.\cite{Heyd2003, Ahlrichs2006}
The error function kernel is used, for instance, in the
Heyd--Scuseria--Ernzerhof (HSE) functionals for solid-state
calculations,\cite{Heyd2003, Heyd2006} as well as the aforementioned
$\omega$B97M-V\cite{Mardirossian2016} functional.
Some functionals based on Yukawa kernels,
$\phi^\text{sr}(r;\omega)=e^{-\lambda r}$, $\phi^\text{lr}(r;\omega)=1
- e^{-\lambda r}$, have also been published and are available in
\LibXC{}, for instance.
It is important to check that the range-separation kernel used in the
density functional implementation matches the one used in the
computation of the range-separated ERIs in \eqref{rseri}.

\subsection{Non-local correlation \label{sec:nlc}}

Dispersion effects, \ie{} van der Waals interactions, can be modeled
in an \emph{ab initio} DFT setting with non-local correlation
functionals\cite{Berland2015}
\begin{equation}
  E^\text{nlc} = \int \dro \drt \rho(\vec{r}_1) \Phi_0 (\vec{r}_1,
  \vec{r}_2) \rho(\vec{r}_2) \label{eq:nlc}
\end{equation}
Because the non-local correlation energy term depends explicitly on
the electron density, it also needs to be included in the SCF
procedure, in principle.
In contrast, empirical dispersion corrections such as Grimme's
various DFT-D approaches\cite{Grimme2004, Grimme2010, Caldeweyher2017}
do not depend on the electron density, and are added only as an
\emph{ad hoc} correction onto the electronic energy.

Perhaps the most accurate rung-3 and rung-4 functionals currently
available,\cite{Najibi2018, Santra2019b, Iron2019} the pure
B97M-V\cite{Mardirossian2015} mGGA as well as the range-separated
$\omega$B97M-V\cite{Mardirossian2016} hybrid mGGA, respectively, are
built on top of\cite{Calbo2015} the VV10 non-local correlation
functional\cite{Vydrov2010} which is controlled by two adjustable
parameters, $b$ and $C$, which are trained alongside the density
functional.
The results of a recent benchmark study suggest that the VV10
contributions on densities and orbital energies are negligible, and
that sufficiently accurate energetics may be obtained by a one-shot
evaluation of $E^\text{nlc}$ in a post-SCF fashion.\cite{Najibi2018}
Still, a rigorous minimization of the energy requires considering the
effects of the non-local correlation on the wave function.
Although \eqref{nlc} does not appear to fit on the rungs of
functionals discussed above, the VV10 kernel turns out to yield a
GGA-type contribution to the Kohn--Sham--Fock matrix as discussed in
\citeref{Vydrov2010}, to which we refer for further details.

\section{Summary and discussion \label{sec:summary}}
We have presented an overview of self-consistent field calculations
within a variational basis set formalism, and discussed the solution
of the self-consistent field equations arising from Hartree--Fock as
well as various levels of density functional approximations using
either the traditional fixed-point equations or direct methods, as
well as various conceptual and numerical issues arising in their
implementation.
No assumptions have been made on the underlying basis set in the
present work: the self-consistent field formalism is the same
regardless of the form of the basis functions, which can be chosen to
be \eg{} Gaussian-type orbitals (GTOs), Slater-type orbitals (STOs),
numerical atomic orbitals (NAOs), or finite element shape functions.
The basis set is only reflected in the molecular integrals, that is,
the matrix elements of the core Hamiltonian and the two-electron
integrals.
The various choices for the basis set have different advantages and
disadvantages, including the evaluation of the molecular integrals;
see \citeref{Lehtola2019c} for discussion.
Instead, the main restriction in the present work is the implicit assumption
that the basis set is compact enough so that the $N \times N$ density
and Fock matrices are small enough to allow $O(N^2)$ dense matrix
storage and $O(N^3)$ diagonalization.
Although the present overview has focused on non-relativistic
calculations on molecules, the discussion for relativistic
calculations as well as crystalline systems is analogous to a large
degree, as 
We hope that the present, consistent and thorough derivation will be
useful for reference as well as teaching purposes, and that the
results presented herein will lead to a wider availability of density
functionals in electronic structure programs.

\section*{Acknowledgments}
This work has been supported by the Academy of Finland (Suomen
Akatemia) through project number 311149.

\bibliography{citations}

\end{document}